# Ab initio Calculations of the Interface States of Polyacetylene-Polyvinylfluoride and Polyethylene - Polyvinylfluoride Quasi-one-dimensional Chains

**Mohamed Assad Abdel-Raouf**
**Physics Department, College of Science, United Arab Emirates,**
**Al-Ain, United Arab Emirates**

## Abstract

The interface states appearing in polyacetylene-polyvinylfluoride and polyethylene-polyvinylfluoride are determined via an ab initio self consistent field technique based on Green's matrix formalism. Different properties of these states are explored. Contrary to the results of the second pair, the results of the first pair showed that the active electronic structure of polyacetylene leads to new states lying in the energy gap of polyvinylfluoride which enhances the doping probability in the first pair. The results emphasize the appearance of bending band phenomenon as a result of the interface of systems considered.

-------------------------------------------------------------------------------------------------

# 1. INTRODUCTION

Polyvinylfluoride (PVF), $(CH_2\text{-}CHF)x$, and its fluorine and chlorine derivatives are polymers possessing very stable physical properties [1]. They have been widely investigated and synthesized for industrial purposes since the early fifties of the preceding century [2]. Their correct x-ray and electron diffraction analyses are known since the work of Bunn and Howell [3] and Clark and Muus [4], respectively. Band structure calculations of these polymers using an ab initio self consistent field (SCF) approach based on linear combination of atomic orbitals (LCAO) were performed by many authors [5]. Earlier investigations of their electronic properties [6] in cis and trans configurations emphasized the argument that PVF and PVDF are better intrinsic semiconductors than polyacetylene (PA) and could be employed for forming high conducting n-doped materials. On the other hand, copolymerization of PVF and PA indicated that the addition of PVF to trance PA does not only increase the intrinsic semiconductivity of PA but also enables easier doping. Moreover, the investigations of the interface problems arising at the interaction surfaces between various polymeric structures have gained tremendous interest in the last decade. This is mainly attributed to their large technological applications in electronic equipments.

The purpose of the present work is to localize the interface states appearing between polyacetylene - polyvinylfluoride (PA-PVF) as well as polyethylene - polyvinylfluoride (PE-PVF) quasi-one-dimensional chains. Special attention is devoted to the nature of mutual interactions arising in each pair. The band structures of the constituents of each pair are determined using the ab initio self consistent field (SCF) approach based on linear combination of atomic orbitals (LCAO). The interface states are located via a Green's matrix formalism employed previously by the author [7]. The main bulk of the present paper is presented in the following two sections. Section 2 contains a brief account on the mathematical formalism, whilst section 3 is devoted to the results and discussions of our calculations. The work ends with a conclusion and the list of references appearing in the text.

## 2. Green's Matrix Formalism

The mathematical formalism suggested for the investigation of the interface problem between two quasi-one-dimensional chains proceeds with the establishment of their band structures using an ab initio self-consistent field (SCF) linear combination of atomic orbitals (LCAO) subjected to the solution of Fock's eigenvalue problem

$$\underline{F}_i(k_\ell)\underline{c}_{i\ell}(k_\ell) = E_{i\ell}(k_\ell)\underline{S}_\ell(k_\ell)\underline{c}_{i\ell}(k_\ell) \qquad \ell = 1,2; i = 1,2,.. \quad (1)$$

Where $\underline{F}_i(k_\ell)$ and $\underline{S}_\ell(k_\ell)$ are the Fock and overlap matrices of the chain i $\ell = 1,2$ .n the momentum space. They can be expressed n the real space by the Fourier transforms

$$F_\ell(k_\ell) = \sum_{q=-\infty}^{q=\infty} e^{ik_\ell q} F_\ell(q) \qquad (2)$$

$$S_\ell(k_\ell) = \sum_{q=-\infty}^{q=\infty} e^{ik_\ell q} S_\ell(q) \qquad (3)$$

$E_{i\ell}(k_\ell)$ and $\underline{c}_{i\ell}(k_\ell)$, eq. (1), identify, respectively, the corresponding eigenvalues and eigenvectors, i.e. the bands energies and wavefunctions ($\psi_{i_\ell}$) , respectively. The latter might be expressed in terms of the crystal orbitals employed in the SCF LCAO calculations, i.e. $\chi^q_{\ell,\nu}$ , as

$$\psi_{i_\ell}(k_\ell, r) = \frac{1}{(N_\ell + 1)^{1/2}} \sum_{q=-N_\ell/2}^{q=N_\ell/2} \sum_{\nu=1}^{m_\ell} e^{ik_\ell q} c_{i_\ell,\nu}(k_\ell) \chi^q_{\ell,\nu} \qquad (4)$$

Where $N_\ell = \infty$ in case of infinite periodic chains, $m_\ell$ is the number of components of the basis set .in the unit cell of chain $\ell$. The elements of the Fock and overlap matrices can be expressed in terms of the basis set components by

$$F_{\ell;\mu,\nu}(q) = F^{q_1,q_2}_{\ell;\mu,\nu} = H^{q_1,q_2}_{\ell;\mu,\nu} + W^{q_1,q_2}_{\ell;\mu,\nu}, \qquad q = q_1 - q_2$$

(5)

$$S^{q_1,q_2}_{\ell;\mu,\nu} = \langle \chi^{q_1}_{\ell,\mu} | \chi^{q_2}_{\ell,\nu} \rangle$$

(6)

The one-electron part of the elements of the Fock matrix is given by

$$H^{q_1,q_2}_{\ell;\mu,\nu} = -\frac{1}{2} \langle \chi^{q_1}_{\ell,\mu} | \chi^{q_2}_{\ell,\nu} \rangle - \sum_{\alpha=-N_\ell/2}^{N_\ell/2} \sum_{A=1}^{M_\ell} \langle \chi^{q_1}_{\ell,\mu} | \frac{Z_A}{|r - R_\alpha - R_A|} | \chi^{q_2}_{\ell,\nu} \rangle \qquad (7)$$

where $M_\ell$ refers to the number of nuclei in the unit cell of chain $\ell$. $Z_A$ is the nuclear charge of the $A^{th}$ atom, $R_\alpha$ is the position vector of a given point in the unit cell. r and $R_A$ are, respectively and the position vectors of the electron and the $A^{th}$ nucleus relative to that point.

The matrix elements representing the two electrons interactions are given by

$$W^{q_1,q_2}_{\ell;\mu,\nu} = \sum_{\mu'=1}^{n_b^\ell} \sum_{\nu'=1}^{n_b^\ell} \sum_{q_1'=-\infty}^{\infty} \sum_{q_2'=-\infty}^{\infty} P^{q_1',q_2'}_{\ell;\mu',\nu'} | \langle \chi^{q_1}_{\ell,\mu}(1) \chi^{q_1'}_{\ell,\mu'}(2) | \frac{1}{r_{12}}(1 - \tfrac{1}{2}\hat{P}_{1\Leftrightarrow 2}) | \chi^{q_2}_{\ell,\nu}(1) \chi^{q_2'}_{\ell,\nu'}(2) \rangle |$$

(8)

where $\hat{P}_{1 \Leftrightarrow 2}$ is the exchange operator. $P_{\ell;\mu',\nu'}^{q_1',q_2'}$ is an element of the so called charge-bond-order matrix. of the chain $\ell$ It is calculated by

$$P_{\ell;\mu',\nu'}^{q_1',q_2'} = \frac{a_\ell}{2\pi} \sum_{i_\ell=1}^{n_{occ}^\ell} \int_{-\pi/a_\ell}^{\pi/a_\ell} e^{ik_\ell(q_1'-q_2')} c_{i_\ell,\mu'}(k_\ell) c_{i_\ell,\nu'}(k_\ell) dk_\ell \qquad (9)$$

The Greek indices characterize the components of the basis set; the Latin indices specify a given cell, they run from the cell $-N_g^\ell$ to the cell $N_g^\ell$; ($N_g^\ell$ is the number of neighbors taken into account). $n_{occ}^\ell$ is the number of occupied bands of chain $\ell$ which coincides with the number of doubly occupied orbitals in the cell. $a_\ell$ denotes the elementary translation parameter of the same chain.

From eqs. (2), (8) and (9), it is obvious that the Fock matrices depend on the eigenvectors $c_{i_\ell}(k_\ell)$, i.e. the problem is nonlinear and should be solved iteratively until self consistency is reached. The Green matrix of the unperturbed $\ell$ chain is defined in terms of its overlap and Fock matrices by

$$\underline{G}_\ell^0(z) = (z\underline{S}_\ell - \underline{F}_\ell)^{-1}, \ell = 1,2, \qquad (10)$$

where $z = E_\ell + i\varepsilon$ and $\varepsilon$ is an infinitesimal positive number. The elements of $\underline{G}_\ell^0(z)$ is given by

$$(\underline{G}_\ell^0(z))_{\mu,q_1,\nu,q_2} = \frac{a_\ell}{2\pi} \sum_{i_\ell}^{n_{occ}^\ell} \int_{-\pi/a_\ell}^{\pi/a_\ell} \frac{c_{i_\ell,\mu}(k_\ell) c_{i_\ell,\nu}(k_\ell) e^{ik_\ell(q_1-q_2)}}{z - E_{i_\ell}(k_\ell)} dk_\ell . \qquad (11)$$

The interaction between the two chains is subjected to Dyson's equation

$$\underline{G}_\ell^{(0)}(E) = \{I - \underline{G}_\ell^0(E)\underline{V}_\ell^{(0)}(E)\}^{-1} \underline{G}_\ell^0(E), \qquad \ell = 1,2 \qquad (12)$$

where I is a unit matrix, $V_\ell^{(0)}$ is the interaction matrix of the corresponding chain which is defined by

$$\underline{V}_\ell^{(0)}(E) = \lim_{\varepsilon \to 0}(E+i\varepsilon)[S_1(k_\ell) - S_2(k_\ell)] - [H_1(k_\ell) - H_2(k_\ell) + \underline{W}_1^{(0)}(k_1) - \underline{W}_2^{(0)}(k_2)]$$

$$= \lim_{\varepsilon \to 0}(E+i\varepsilon)\Delta S_\ell(k_\ell) - \Delta F_\ell^{(0)}(k_\ell) \tag{13}$$

Such that $\underline{V}_2^{(0)}(E) = -\underline{V}_1^{(0)}(E)$. (14)

Remember that the Green matrix $\underline{G}_\ell^{(0)}(E)$ is nonsingular outside the bands of the chain. Consequently, the poles of this matrix define the interface states (IFS) arising through the mutual interaction between the chains; they are identical with the zeros of the determinate equation

$$\underline{D}_\ell^{(0)}(E) = \det\{I - \underline{G}_\ell^0(E)\underline{V}_\ell^{(0)}(E)\}^{-1}, \qquad \ell = 1,2 \tag{15}$$

These roots are confined by the variational principle such that

$$E_b^{\ell(0)} \leq E_{b+1}^{\ell(0)}, \quad b = 1,2,3,.. \ , \ \ell = 1,2. \tag{16}$$

Remark that the roots of eq. (15) depend on the quality of the interaction matrix $\underline{V}_\ell^{(0)}(E)$ delivered by the SCF LCAO calculations. Thus, they can be improved self consistently by using eq. (12) for determining a new charge-bond-order matrices for the two chains, new Fock matrices via eqs.(5), (7) - (9) and new interaction matrices by employing eqs. (13) and (14). The new charge-bond-order matrices are obtained by

$$\underline{P}_\ell^{(1)} = -\frac{1}{\pi}\int_{E^\ell_L}^{E^\ell_U} \text{Im}[\underline{G}_\ell^{(0)}(E)]\, dE, \tag{17}$$

Where $E^\ell_L$ and $E^\ell_U$ are, respectively, the lowest and upper edges of the highest occupied band.. Trustworthy results could be obtained iteratively by repeating the preceding steps until the convergence has been reached. On the other hand, a valuable indication for the existence of the interface states is given by Callaway's phase shifts [8] $\delta_\ell^{(\upsilon)}(E)$, where $\upsilon$ is the order of iteration. These are related to the deviation, $\Delta\rho_\ell^\upsilon(E)$, of the density of states of chain $\ell$ from its unperturbed one by

$$\Delta\rho_\ell^{(\nu)}(E) = \frac{1}{\pi}\frac{d\delta_\ell^{(\nu)}(E)}{d\varepsilon} = \frac{1}{\pi}\frac{d}{d\varepsilon}\left[-\tan^{-1}\left[\frac{\text{Im}[D_\ell^{(\nu)}(E)]}{\text{Re}[D_\ell^{(\nu)}(E)]}\right]\right] \quad (18)$$

The cases $\text{Re}[D_\ell^{(\nu)}(E)] = 0$ are anomalies at which Callaway's phases are odd multiple of π/2. They identify the interface states falling in the bands of the unperturbed chain, the total density of states is determined by

$$\rho_\ell^{(\nu)}(E) = -\frac{1}{2\pi}Tr\left[\text{Im}(\underline{G}_\ell^{(\nu)}(E)\underline{S}_\ell + \underline{S}_\ell\underline{G}_\ell^{(\nu)})\right]. \quad (19)$$

3. Results and Discussion

As mentioned in the preceding section the first step towards the localization of the interface states occurring in the two pairs: trans polyacetylene – polyvinylfluoride and polyethylene - polyvinylfluoride is the performance of the accurate ab initio SCF LCAO calculations of the band structures of the three involved chains. This was accomplished by employing the standard geometry [11]: C-C = 1.36 $A^0$, C-H = 1.08 $A^0$, translation distance = 2.41 $A^0$ for PA; C – C = 1.54 $A^0$, C – H = 1.09 $A^0$, translation distance = 2.52 $A^0$ for PE and C – C = 1.54 $A^0$, C – H = 1.09 $A^0$, C – F = 1.32 $A^0$ and translation distance = 2.52 $A^0$ for PVF. The angle HCH is equal to 109.5°. The minimal basis sets developed by Hehre, Stewart and Pople [12] were employed with 12 components for PA, 14 for PE and 18 for PVF. The number of occupied bands ($n^\ell_{occ}$) is 7 for PA, 8 for PE and 11 for PVF. The band structure of PA and PE were determined using 6 neighbors (i.e. Ng = 6) interaction approximation, whilst the first neighbor interaction approximation was employed for the investigation of the band structure of PVF.

Table 1 contains the values of the lower and upper edges (in a.u) of the occupied bands of PA, PE and PVF based on the preceding approximations. The Table indicates the energy domains employed in the calculation of the charge-bond-order matrices, (see eq. (17)), appearing in the present interface problems (i.e. PA-PVF and PE-PVF). Thus, $E_L$ = -26.239 a.u., $E_U$ = 0.292 were used in the first system, while $E_L$ = - 26.239 a.u. and $E_U$ = -0.451 a.u. were used in the second system.

The investigations of the band structure were followed by the determination of the Green matrices of the unperturbed chains using eq. (11). The integrals appearing in this equation were carried out via Simpson's rule with 9 mesh points.

The overlap and Fock matrices of the chains were used for determining the interactions matrices of the pairs via eqs. (13) and (14), the Green matrices of the perturbed chains via eq. (12) and the real and imaginary parts of the determinants presented at eq. (15). The imaginary parts of the perturbed Green matrices were employed for evaluating new charge-bond-order matrices via eq. (17) with the help of Simpson's expansions of the integrals in terms of 181 mesh points. The interface states were identified as the roots of the real parts of eq. (15). These roots were obtained by a least-squares fitting of the real parts with a polynomial of degree 19. In Table 2 are accumulated the resulting interface states of the PA-PVF system. Comparisons between Tables 1 and 2 yield the following remarks:

(1)The first 10 IFS of PA lie below its first occupied band, states no. 11-16 lie in the gap between the second and third bands, state no 17 lies in fourth band, state no 18 lies in the seventh band and state no 19 occurs above the valence band of the PA chain. Thus, existence of PVF increases the doping probability of the PA chain.

(2) The first 10 IFS of PVF lie inside the gap between the first and second occupied bands, states nos. 11-16 lie between the third and fourth bands, state no. 17 lies in the gap between the fourth and fifth bands, state no 18 occurs between the bands nos. 8 and 9 and the last IFS lies in the energy gap of PVF. In other words, the existence of PA enhances the doping probability of the PVF chain.

(3) The last IFS of the PA - PVF system support the argument the copolymerization of PA and PVF enhances the doping probability of its constituents.

The interface states of the PE-PVF system are given in Table 3. The results emphasize the following points:

(i) The first 10 IFS of PE lie below its first occupied band, the IFS nos. 11-17 occur in the gap between the second and third bands, state no. 18 lies in the third band, and the last state exists in the fourth band of the chain. This indicates that the electric property of PE remains very stable, inspite of appearance of new IFS.

(ii) The first 10 IFS of PVF lie inside the gap between its second and third occupied bands, IFS nos. 11-16 lie between the third and fourth bands, state no. 17 shows up between the fourth and fifth bands, state no. 18 lies between the sixth and seventh bands and the last interface state lies in the last band. This means that no IFS appear in the energy gap of PVF and, consequently, no improvements upon the electric property of PVF through PE and vice versa.

On comparing the last columns of Tables 2 and 3 as well as points (2) and (ii) we realize that the active electronic structure of the polyacetylene chain plays an important role in the appearance of the interface states between PA and PVF. On the contrary, the saturated property of PE prevents the appearance of interface states within the energy gaps of PVF. Also from Tables 2 and 3, one concludes that the interface interaction between each pair of the investigated chains smooth out the distribution of states inside the sea of occupied bands and, consequently, lead to the appearance of the bending bands phenomena inside its constituents.

**Table 1: Lower and upper edges (in a.u.) of the occupied bands of polyacetylene (PA), polyethylene (PE) and polyvinylfluoride (PVF) chains.**

| Band order | PA | | PE | | PVF | |
|---|---|---|---|---|---|---|
| | Lower | Upper | Lower | Upper | Lower | Upper |
| 1 | -11.4498 | -11.4493 | -11.3972 | -11.3972 | <u>- 26.2388</u> | -26.2388 |
| 2 | -11.4492 | -11.4488 | -11.3970 | -1103969 | -11.5130 | -11.5128 |
| 3 | -1.2146 | -0.9998 | -1.1739 | -0.9654 | -11.4255 | -11.4252 |
| 4 | -0.9745 | -0.8042 | -0.9604 | -0.8280 | -1.6140 | -1.5908 |
| 5 | -0.7550 | -0.6827 | -0.6674 | -0.6299 | -1.1711 | -1.0152 |
| 6 | -0.6743 | -0.5897 | -0.6284 | -0.6106 | -0.9719 | -0.8556 |
| 7 | -0.5837 | <u>-0.2920</u> | -0.6044 | -0.5857 | -0.7718 | -0.7523 |
| 8 | | | -0.5839 | -0.5211 | -0.7412 | -0.7254 |
| 9 | | | | | -0.6884 | -0.6874 |
| 10 | | | | | -0.6217 | -0.6165 |
| 11 | | | | | -0.5752 | -0.4949 |
| 12 | | | | | -0.5212 | <u>-0.4510</u> |

**Table 2:** The interface states (in a.u.) appearing in the chains of the PA-PVF system. (*) assign states lying in the energy gap of the chain. The underlined state occurs inside an occupied band.

| Order of the inter face state | PA | PVF |
|---|---|---|
| 1 | -26.206 | -26.039 |
| 2 | -25.793 | -25.729 |
| 3 | -25.154 | -24.981 |
| 4 | -24.023 | -24.019 |
| 5 | -22.713 | -22.646 |
| 6 | -21.076 | -21.124 |
| 7 | -19.244 | -19.311 |
| 8 | -17.230 | -17.402 |
| 9 | -15.075 | -15.306 |
| 10 | -12.891 | -13.183 |
| 11 | -10.705 | -11.058 |
| 12 | -8.570 | -9.017 |
| 13 | -6.559 | -7.027 |
| 14 | -4.738 | -5.258 |
| 15 | -3.151 | -3.698 |
| 16 | -1.859 | -2.393 |
| 17 | <u>-0.934</u> | -1.380 |
| 18 | <u>-0.347</u> | -0.691 |
| 19 | 1.947 | 0.335 (*) |

**Table 3: The interface states (in a.u.) appearing in the chains of the PE-PVF system. The underlined state occurs inside an occupied band.**

| Order of the inter face state | PE | PVF |
|---|---|---|
| 1 | -26.204 | -26.065 |
| 2 | -25.852 | -25.667 |
| 3 | -25.127 | -25.050 |
| 4 | -24.141 | -23.990 |
| 5 | -22.827 | -22.703 |
| 6 | -21.268 | -21.140 |
| 7 | -19.514 | -19.373 |
| 8 | -17.553 | -17.432 |
| 9 | -15.525 | -15.376 |
| 10 | -13.407 | -13.270 |
| 11 | -11.303 | -11.164 |
| 12 | -9.225 | -10.110 |
| 13 | -7.305 | -7.171 |
| 14 | -5.537 | -5.396 |
| 15 | -3.996 | -3.841 |
| 16 | -2.681 | -2.542 |
| 17 | <u>-1.662</u> | -1.534 |
| 18 | <u>-0.985</u> | -0.849 |
| 19 | <u>-0.634</u> | -0.494 |